\documentclass[journal]{IEEEtran}
\usepackage{amsmath}
\allowdisplaybreaks[4]

\usepackage[justification=centering]{caption}

\usepackage{array}
\usepackage{graphics} 
\usepackage{graphicx}
\usepackage{epsfig} 
\usepackage{times} 
\usepackage{amsmath} 
\usepackage{amssymb}  

\newtheorem{remark}{Remark}
\newtheorem{assumption}{Assumption}

\newtheorem{lemma}{Lemma}
\newtheorem{definition}{Definition}
\newtheorem{theorem}{Theorem}
\newtheorem{proposition}{Proposition}
\ifCLASSINFOpdf

\else

\fi

\begin{document}

\title{New Class $\mathcal{K}_\infty$ Function-Based Adaptive Sliding Mode Control Design}

\author{Jiawei~Song,
        Zongyu~Zuo,~\IEEEmembership{Senior Member}
        Michael~Basin,~\IEEEmembership{Senior Member}
\thanks{J.~Song and Z.~Zuo are with The Seventh Research Division, Beihang University (BUAA), Beijing 100191, China (jeforis@163.com, zzybobby@buaa.edu.cn).}
\thanks{M. Basin is with The Department of Physical and Mathematical Sciences, Autonomous University of Nuevo Leon, San Nicolas de los Garza 66450, Mexico (mbasin@fcfm.uanl.mx)}}

\markboth{IEEE Transactions on Automatic Control}%
{Shell \MakeLowercase{\textit{et al.}}: Bare Demo of IEEEtran.cls for IEEE Journals}

\maketitle

\begin{abstract}
To reduce the chattering and overestimation phenomena existing in classical adaptive sliding mode control, this paper presents a new class $\mathcal{K}_\infty$ function-based adaptive sliding mode control scheme. Two controllers are proposed in terms of concave and convex barrier functions to implement this kind of control methodology. To avoid large initial control magnitudes, two modified control schemes are provided, which extend the proposed methodology to different scenarios. It is proven that the proposed controllers yield finite-time convergence to a real sliding mode. Finally, simulations and discussions are presented to show the advantages and effectiveness of the proposed control methodology.
\end{abstract}


\IEEEpeerreviewmaketitle

\section{Introduction}

\IEEEPARstart{T}{o} improve robustness of a closed-loop system, sliding mode control is always a good choice. It has been put into practice in numerous application fields. To achieve finite-time convergence \cite{Bhat 2000} to the sliding manifold, a discontinuous control is conventionally designed, which however causes one of the well-known problems, chattering. The chattering is not welcome in engineering applications, since it can burden the actuators and, more seriously, the system may break down in a short time. Nowadays, a variety of methods for attenuating the chattering have been reported.

Introducing a boundary layer \cite{Slotine 1991}, \cite{Zhu 2011} is one of the commonly used adaptive techniques in the sliding mode control. The saturation function can be used to approximate the signum function, so the chattering can be reduced to some extent at the price of losing accuracy and robustness \cite{Ding 2013}. Another effective way, called the equivalent control \cite{Utkin 1996}, \cite{Lee 2007}, is also utilized to reduce the chattering by adding a low-pass filter to the control input. Unfortunately, as mentioned in \cite{Plestan 2010}, the low-pass filter can present drawbacks like signal magnitude attenuation, delay, and transient behaviour. To achieve better performance, the time constant of the low-pass filter should be carefully tuned. There are a number of adaptive chattering-suppressing techniques apart of those methods, however, the chattering problem still remains a great difficulty in the sliding mode control field.

To enhance applicability of the sliding mode control, the control constraint due to limited energy or actuator capability should be considered. Note that the invariance with respect to (matched) disturbances is a key feature of the sliding mode control. For a disturbed system, achieving the invariance requires a sufficient control force, that is, the control magnitude should exceed a disturbance bound. Nevertheless, in most cases, disturbance bounds are unknown. An excessive control gain yields a faster convergence rate but causes a larger chattering amplitude. Therefore, determining an appropriate control magnitude matters a lot. To obtain a proper control gain, Huang et al. in \cite{Huang 2008} proposed an adaptive sliding mode control scheme. In \cite{Huang 2008}, the adaptation law was established as a non-decreasing function of the dynamic errors, which leads to overestimation even if the disturbance vanishes. Furthermore, the adaptive gain increases forever due to computational errors and noises. Similar adaptive sliding mode techniques have been used for tracking a flexible air-breathing vehicle \cite{Hu 2012} and stabilizing a networked control system with time-varying delays \cite{Khanesar 2015}. To overcome the deficiencies of this kind of adaptive sliding mode control design, the paper \cite{Plestan 2010} proposed a new adaptation law based on a real sliding mode (instead of an ideal one), which is easier to achieve in a practical system. However, the chattering still exists and the real sliding mode bound needs to be tuned and selected cautiously. If the bound is too small, stability cannot be guaranteed and, otherwise, if too large, accuracy is deteriorated. Following a similar idea of achieving a real sliding mode, the work \cite{Obeid 2018} proposed a barrier function-based adaptive sliding mode control approach. Different from \cite{Plestan 2010}, the adaptation process is divided into two phases. The adaptation law in the reaching phase is given by $\dot{K}=\bar{K}|s|$. Once the real sliding mode is reached, the adaptation gain is governed by a barrier function of the sliding variable. This approach guarantees the finite-time convergence to a predefined real sliding mode without overestimation. To eliminate the reaching phase, the barrier functions have been applied to the adaptive integral sliding mode control \cite{Obeid 2018b}.

Motivated by the preceding discussion, in this paper, new class $\mathcal{K}_\infty$ function-based adaptive sliding mode control design is presented in Section \ref{Sec Motivation} to enhance the approach proposed in \cite{Obeid 2018}, \cite{Obeid 2018b}. The designed control methodology provides the following benefits: i) the barrier functions are taken from the class $\mathcal{K}_\infty$; ii) the real sliding mode bound can be reduced to an arbitrarily small one; iii) no overestimation exists for control gains; iv) the chattering is mitigated. Within the framework of the general class $\mathcal{K}_\infty$ gain design given in Section \ref{section K infty}, concave and convex barrier function-based adaptive sliding mode controllers are proposed and their convergence rates are compared qualitatively in Section \ref{cases for nbf}. Additionally, in Section \ref{sec_methods_avoid}, two barrier function-based adaptive sliding mode control are developed to regulate the reaching phase and avoid a large initial control magnitude.

\section{Preliminaries}
In this section, some definitions and lemmas are recalled. Consider the following first-order system
\begin{align}\label{first-order-system}
  \dot{s}=u+d, s(0)=s_0,
\end{align}
where $s$ is the sliding variable, $u$ is the control input and $|d|\leq \bar{d}$ is a disturbance with an unknown postive bound $\bar{d}$.
\begin{definition}\cite{Levant 1993}\label{ideal sliding surface}
  Given the sliding variable $s$, the `ideal sliding surface' associated with (\ref{first-order-system}) is defined as
  \begin{align}
    S^i=\{s\in \mathbb{R} | |s|=0 \}.
  \end{align}
\end{definition}

\begin{definition}\cite{Levant 1993}\label{real sliding surface}
  Given the sliding variable $s$, the `real sliding surface' associated with (\ref{first-order-system}) is defined as (with $\sigma>0 $)
  \begin{align}
    S^r=\{s\in \mathbb{R} | |s|\leq \sigma \}.
  \end{align}
\end{definition}

\begin{definition}\cite{Khalil}\label{K function}
A continuous function $\alpha:[0,a)\rightarrow [0,\infty)$ is said to belong to class $\mathcal{K}$ if it is strictly increasing and $\alpha(0)=0$. It is said to belong to class $\mathcal{K}_\infty$ if $a=\infty$ and $\alpha(r)\rightarrow \infty$ as $r\rightarrow \infty$.
\end{definition}

\begin{lemma}\cite{Khalil}\label{lemma_k_function}
Let $\alpha_1$ be a class $\mathcal{K}$ functions on $[0,a)$ and $\alpha_2$ be a class $\mathcal{K}_\infty$ function. Denote the inverse of $\alpha_i$ by $\alpha_i^{-1}$, $i=1,2$. Then,
\begin{itemize}
  \item $\alpha_1^{-1}$ is defined on $[0,\alpha_1(a))$ and belongs to class $\mathcal{K}$;
  \item $\alpha_2^{-1}$ is defined on $[0,\infty)$ and belongs to class $\mathcal{K}_\infty$.
\end{itemize}
\end{lemma}

\begin{lemma}\cite{Rockafellar 1970}\label{def cc and cv}
A twice-differentiable function $f$ of a single variable defined on the interval $I$ is
\begin{enumerate}
  \item concave if and only if $\ddot{f}(x)\leq 0$ for all $x$ in the interior of $I$;
  \item convex if and only if $\ddot{f}(x)\geq 0$ for all $x$ in the interior of $I$.
\end{enumerate}
\end{lemma}

\section{Motivation}\label{Sec Motivation}
In this section, an existing result on barrier function-based adaptive sliding mode control is recalled and discussed. .
\begin{lemma}\cite{Obeid 2018}\label{lemma_barrier}
  Given the system (\ref{first-order-system}) with a bounded disturbance $d$, consider the controller
  \begin{align}\label{origin-adaptive-ctrl}
  u=-K(t,s)\text{sign}(s),
  \end{align}
   with an adaptive control gain $K(t,s)$
  \begin{align}\label{origin-adaptive-law}
    K(t,s)=\begin{cases}
    K_a(t), \dot{K}_a=\bar{K}|s|,~~ &\text{if} ~~0<t\leq \bar{t},\\
    K_{\mathrm{psb}}(s), ~~ &\text{if} ~~t> \bar{t},\\
    \end{cases}
  \end{align}
  where $\bar{K}$ is any positive constant, $K_{\mathrm{psb}}(s)=\frac{|s|}{\varepsilon-|s|}$ is a positive semi-definite barrier function.
  Then, for any $s(0)$ and $\varepsilon$, there exists $\bar{t}$, the smallest root of equation $|s|\leq\frac{\varepsilon}{2}$, such that for all $t\geq \bar{t}$ the inequality $|s(t)|\leq \varepsilon$ holds.
\end{lemma}

The barrier function-based adaptive sliding mode control scheme in Lemma \ref{lemma_barrier} presents the following benefits: i) the control input ensures that the sliding variable $s$ reaches a real sliding surface without overestimation of the gain; ii) $\sigma$ in Definition \ref{real sliding surface} can be predefined independently of the disturbance bound. However, several drawbacks are also introduced when applying Lemma \ref{lemma_barrier}: i) because of the switching of $K$ at time instant $\bar{t}$, the control input presents discontinuity; ii) the stability of the system may break down under a sudden change of $|s|>\varepsilon$ due to input saturation or a sudden change of the sliding variable; iii) $\bar{t}$ has to be found by trial and error online.

\begin{figure}
  \centering
  \includegraphics[scale=0.5]{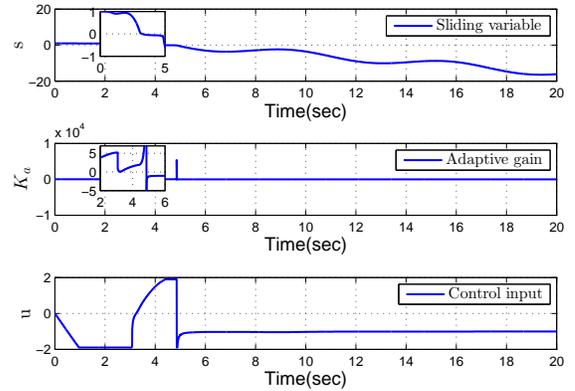}
  \caption{Responses to controller (\ref{origin-adaptive-ctrl}) in presence of input saturation}\label{Saturate impact}
\end{figure}

To show these drawbacks intuitively, the closed-loop responses of the system (\ref{first-order-system}) to the controller (\ref{origin-adaptive-ctrl}) are displayed in Fig. \ref{Saturate impact}. In the numerical simulation, a sinusoidal disturbance $d(t)=2\sin(t)$ is considered. The parameters in (\ref{origin-adaptive-law}) are chosen as $\bar{K}=2$ and $\varepsilon=0.1$. The initial condition $s_0=1$ is assigned. The control input is constrained by $|u|\leq 1.9$. From Fig. \ref{Saturate impact}, one can see the switch of $K$ at $t\approx 3~s$, which presents a discontinuity. Furthermore, the sliding variable escapes from the real sliding surface $|s|\leq 0.1$ after $t>5~s$ and never returns due to the input constraint, which implies a breakdown in the system stability.

\begin{remark} \label{remark motivation}
  Note that, the barrier functions $K_{\mathrm{psb}}(|s|)$ is defined on $[0,\varepsilon)$ and belongs to the class $\mathcal{K}$. This implies that the inverse of $K_{\mathrm{psb}}(|s|)$ also belongs to the class $\mathcal{K}$ by Lemma \ref{lemma_k_function}. Hence, the sliding variable $s$ does not escape from $[0,\varepsilon)$ if and only if the inverse function is defined on $[0,K_{\mathrm{psb}}(\varepsilon))$. However, the positive definiteness of $K(t,s)$ is no longer guaranteed outside the domain $[0,\varepsilon)$ for $t>\bar t$, which implies that the sliding variable never returns to this domain once escaping from it. In practice, the escape may occur under a sudden change of sliding variables, like switching of flight segments in flight control, when an actuator cannot suppress a disturbance at some time interval due to saturation. Hence, in \cite{Obeid 2018}, the control law (\ref{origin-adaptive-ctrl}), (\ref{origin-adaptive-law}) requires an assumption that the maximum control magnitude should be superior to the disturbance bound at every time moment. This condition can be relaxed by using the control technique proposed in this paper, as shown in Fig. \ref{Saturate new_barr} (please see the first example in Section \ref{Sec sim}, based on the control design in Theorems \ref{theorem new barrier-function adaptive} and \ref{thm_Kcv}).
\end{remark}

\section{Main results} \label{section K infty}
To overcome those drawbacks, a novel $\mathcal{K}_\infty$ function-based adaptive sliding mode controller is proposed as:
\begin{align} \label{new_bf_control}
u=-K_{\mathrm{bf}}(|s|)\text{sign}(s),
\end{align}
where $K_{\mathrm{bf}}(|s|)$ is a class $\mathcal{K}_\infty$ function with the following generalized properties:
\begin{description}
\item[P1] $K_{\mathrm{bf}}(|s|)$ defined on $[0,\infty)$ is continuous;
\item[P2] $K_{\mathrm{bf}}(|s|)$ is at least $\mathcal{C}^1$ with respect to $|s|$ on $(0,\infty)$;
\item[P3] The inverse function of $K_{\mathrm{bf}}(|s|)$ exists.
\end{description}

\begin{theorem}\label{theorem new barrier-function adaptive}
Consider the system (\ref{first-order-system}) with the controller (\ref{new_bf_control}). For any $s(0)$, the real sliding surface $|s(t)|\leq \sigma_1$ is reached in finite time, where $\sigma_1 = \kappa_{\mathrm{bf}}(\bar{d})$, $\kappa_{\mathrm{bf}}$ is the inverse of function $K_{\mathrm{bf}}$.
\end{theorem}

\begin{IEEEproof}
By P1, one has $K_{\mathrm{bf}}(|s|)\rightarrow +\infty$ as $|s|\rightarrow +\infty$, therefore, $\dot{K}_{\mathrm{bf}}=\frac{\partial K_{\mathrm{bf}}}{\partial |s|}>0$. For notation simplicity, let $K_{\mathrm{bf}}:=K_{\mathrm{bf}}(|s|)$. Choose the Lyapunov function as $V=\frac{1}{2}s^2$. The full derivative of $V$ in time is calculated as
\begin{align}\label{time-derivative of V1}
\dot{V}=&s(-K_{\mathrm{bf}}\text{sign}(s)+d)\notag\\
=&\big(-K_{\mathrm{bf}}|s|+d s\big)\notag\\
\leq& (-K_{\mathrm{bf}}+\bar{d})|s|\notag\\
\leq& - \sqrt{2}\beta V^{\frac{1}{2}}, \beta=(K_{\mathrm{bf}}-\bar{d})>0.
\end{align}
Let $\sigma_1=\kappa_{\mathrm{bf}}(\bar{d})$, where $\kappa_{\mathrm{bf}}$ is the inverse of $K_\mathrm{bf}$. It follows from Lemma \ref{lemma_k_function} that $\kappa_{\mathrm{bf}}$ is also a class $\mathcal{K}_\infty$ function. Hence, if $|s|>\sigma_1$, one has $K_{\mathrm{bf}}>\bar{d}$. Thus, the finite-time convergence to $|s|\leq \sigma_1$ is proven, where $\sigma_1$ is finite for a given upper-bound $\bar{d}$.
\end{IEEEproof}
\begin{remark}
In practical applications, $K_{\mathrm{bf}}$ should be carefully designed such that the ultimate bound $\sigma_1$ associated with $\kappa_{\mathrm{bf}}$ could be reduced.
\end{remark}

\begin{remark}
Different from \cite{Obeid 2018}, the proposed class $\mathcal{K}_\infty$ barrier function approach guarantees the global finite-time convergence to a real sliding surface. In other words, if an escape from a real sliding mode occurs at a certain time instant, a sliding variable can return to the real sliding mode once $K_{\mathrm{bf}}>|d(t)|$, as shown in Fig. \ref{Saturate new_barr}.
\end{remark}

\section{Controller Design}\label{cases for nbf}
Based on Theorem \ref{theorem new barrier-function adaptive}, two novel sliding mode controllers associated with two different barrier functions belonging to class $\mathcal{K}_\infty$, including a concave function and a convex function, are developed in this section.
\subsection{Concave barrier function-based adaptive sliding mode controller}\label{subsec_concave}
\begin{figure}
  \centering
  \includegraphics[scale=0.45]{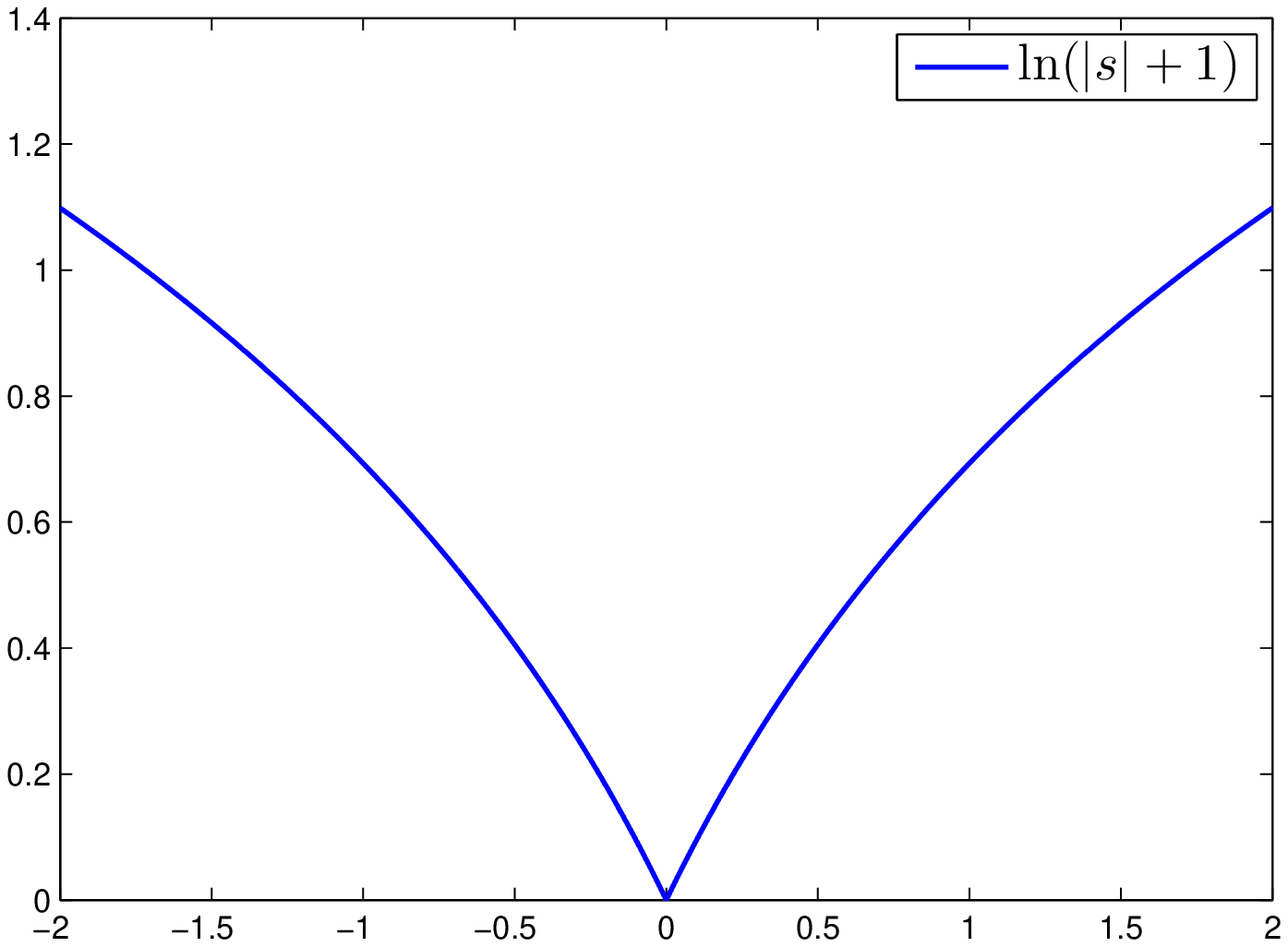}
  \caption{$K_{\mathrm{cc}}=\ln(|s|+1)$}\label{fig-kcc}
\end{figure}

In this subsection, the following concave barrier function is considered:
\begin{align}
K_{\mathrm{cc}}(|s|)=\rho_1\ln\left(\frac{|s|}{\lambda_1}+1\right),
\end{align}
where $\rho_1$ and $\lambda_1$ are positive parameters. By Lemma \ref{def cc and cv}, $K_{\mathrm{cc}}(\cdot)$ is a strictly increasing concave function (see Fig. \ref{fig-kcc}) on $(0,+\infty)$, since its first-order derivative ${K}^{'}_{\mathrm{cc}}(|s|)=\frac{\rho_1}{\lambda_1+|s|}>0$ and the second-order derivative ${K}^{''}_{\mathrm{cc}}(|s|)=-\frac{\rho_1}{(\lambda_1+|s|)^2}<0$ on $(0,+\infty)$.

\begin{theorem}\label{thm_Kcc}
Consider the system (\ref{first-order-system}) with the following controller:
\begin{align} \label{cc-barrier}
u=-K_{\mathrm{cc}}(|s|)\mathrm{sign}(s).
\end{align}
Then, for any $s_0$, the real sliding surface $|s(t)|\leq \sigma_{\mathrm{cc}}$ is reached in a finite time, where $\sigma_{\mathrm{cc}}$ can be reduced to arbitrarily small by tuning $\rho_1$ and $\lambda_1$.
\end{theorem}

\begin{IEEEproof}
Obviously, properties P1-P3 in Theorem \ref{theorem new barrier-function adaptive} hold for $K_{\mathrm{cc}}$. Likewise, choose the Lyapunov function $V=\frac{1}{2}s^2$, then the full time derivative of $V$ is given by:
\begin{align}\label{dVcc}
\dot{V}&=s(-K_{\mathrm{cc}}\text{sign}(s)+d)\notag\\
&\leq (-K_{\mathrm{cc}}+\bar{d})|s|\notag\\
&\leq -\beta_{\mathrm{cc}} V^{\frac{1}{2}}, \beta_{\mathrm{cc}}=(K_{\mathrm{cc}}-\bar{d})\sqrt{2}>0.
\end{align}
This ensures the finite-time convergence of $s$ to the real sliding surface. The following ultimate bound can be calculated:
\begin{align}\label{kappa_cc_d}
\sigma_{\mathrm{cc}}=\lambda_1\left(e^{\frac{\bar{d}}{\rho_1}}-1\right),
\end{align}
which shows that $\sigma_{\mathrm{cc}}$ can be reduced to arbitrarily small by increasing $\rho_1$ or decreasing $\lambda_1$.
\end{IEEEproof}

\subsection{Convex barrier function-based adaptive sliding mode controller}
\begin{figure}
  \centering
  \includegraphics[scale=0.45]{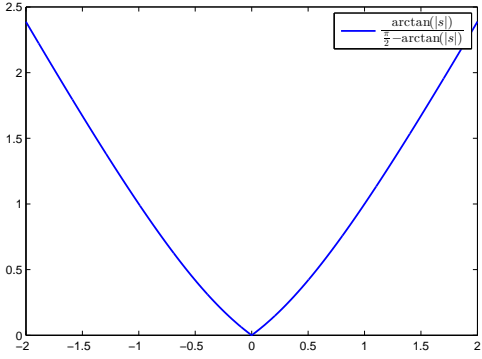}
  \caption{$K_{\mathrm{cv}}=\frac{\arctan(|s|)}{\frac{\pi}{2}-\arctan(|s|)}$}\label{fig-kcv}
\end{figure}

Likewise, the following convex function is considered:
\begin{align}
K_{\mathrm{cv}}(|s|)=\rho_2\frac{\arctan(\frac{|s|}{\lambda_2})}{\frac{\pi}{2}-\arctan(\frac{|s|}{\lambda_2})},
\end{align}
where $\rho_2$ and $\lambda_2$ are positive parameters. By Lemma \ref{def cc and cv}, $K_{\mathrm{cv}}(\cdot)$ is a strictly increasing convex function (see Fig. \ref{fig-kcv}) on $(0,+\infty)$ since
\begin{align}\notag
K^{'}_{\mathrm{cv}}(|s|)=\frac{2\rho_2\lambda_2\pi}{\left(\pi-2\arctan(\frac{|s|}{\lambda_2})\right)^2(\lambda_2^2+s^2)}>0,
\end{align}
and
\begin{align}\notag
 K^{''}_{\mathrm{cv}}(|s|)=\frac{4\pi\rho_2\lambda_2\left(2\lambda_2-\pi|s|+2|s|\arctan(\frac{|s|}{\lambda_2})\right)}{\left(\pi-2\arctan(\frac{|s|}{\lambda_2})\right)^3(\lambda_2+s^2)^2}>0,
\end{align}
on $(0,+\infty)$.
\begin{theorem}\label{thm_Kcv}
Consider the system (\ref{first-order-system}) with the following controller:
\begin{align} \label{cv-barrier}
u=-K_{\mathrm{cv}}(|s|)\mathrm{sign}(s).
\end{align}
Then, for any $s_0$, the real sliding surface $|s(t)|\leq \sigma_{\mathrm{cv}}$ is reached in a finite time, where $\sigma_{\mathrm{cv}}$ can be reduced to arbitrarily small by tuning $\rho_2$ and $\lambda_2$.
\end{theorem}

\begin{IEEEproof}
Following the same lines as the proof of Theorem \ref{thm_Kcc} with
\begin{align}
\beta_{\mathrm{cv}}=(K_{\mathrm{cv}}-\bar{d})\sqrt{2}>0
\end{align}
and
\begin{align}\label{kappa_cv_d}
\sigma_{\mathrm{cv}}=\lambda_2\tan\left(\frac{\pi\bar{d}}{2(\bar{d}+\rho_2)}\right)
\end{align}
implies that $\sigma_{\mathrm{cv}}<\infty$, i.e., $\sigma_{\mathrm{cv}}$ is bounded. To reduce the bound $\sigma_{\mathrm{cv}}$ to arbitrarily small, $\rho_2$ can be chosen sufficiently large or $\lambda_2$ can be selected sufficiently small.
\end{IEEEproof}

\subsection{Comparison of convergence times}
For the sake of fairness, the closed-loop dynamics governed by (\ref{cc-barrier}) and (\ref{cv-barrier}) are considered with the same convergence accuracy $|s|\leq \sigma_c$ and the same initial condition $|s_0|>\sigma_c$. The comparison of the on convergence times is given by the following proposition.

\begin{proposition}\label{prop comp}
Given the same convergence accuracy and initial condition, the closed-loop system (\ref{first-order-system}) with (\ref{cv-barrier}) yields a faster convergence rate than the one with (\ref{cc-barrier}).
\end{proposition}
\begin{IEEEproof}
Consider the Lyapunov function $V=|s|$. It follows that
\begin{align}\label{comp_Lyap}
  \dot{V}\leq -\mu(s(t)),
\end{align}
where $\mu(t)=K_{\mathrm{bf}}(|s(t)|)-\bar{d}$ with $K_{\mathrm{bf}}(|s(t_0)|)=K_{\mathrm{bf}}(|s_0|)$ and $K_{\mathrm{bf}}(|s(t_f)|)=K_{\mathrm{bf}}(\sigma_c)$, where $t_f$ is the convergence time. The functions $K_{\mathrm{cc}}$ and $K_{\mathrm{cv}}$ intersect at two points, $|s|=0$ and $|s|=\sigma_c$. Considering the fact that $K_{\mathrm{cc}}$ is concave and $K_{\mathrm{cv}}$ is convex, one has $K_{\mathrm{cc}}>K_{\mathrm{cv}}$ for $|s|\in(0,\sigma_c)$ and $K_{\mathrm{cc}}<K_{\mathrm{cv}}$ for $|s|\in (\sigma_c,+\infty)$. From (\ref{comp_Lyap}), it can be concluded that the closed-loop system (\ref{first-order-system}) with (\ref{cv-barrier}) yields a faster convergence rate than that with (\ref{cc-barrier}).
\end{IEEEproof}

\section{Further Extensions}\label{sec_methods_avoid}
Section \ref{cases for nbf} presents two different barrier function-based adaptive sliding mode controllers which guarantee finite-time convergence to a real sliding mode. However, it should be noted that a high accuracy may lead to a large initial control magnitude, especially in the concave function case. This section provides two possible solutions to this problem.
\subsection{Saturated barrier function-based adaptive sliding mode control}
To avoid a large initial control magnitude, a direct solution is to impose a constraint on the control input. Then, the following saturation function is considered:
\begin{align}\label{sat}
  \text{sat}(K_{\mathrm{bf}}(t,s),\bar{K}_{\mathrm{bf}})=\begin{cases}
  \bar{K}_{\mathrm{bf}},~~&\text{if}~~ K_{\mathrm{bf}}\geq \bar{K}_{\mathrm{bf}},\\
  K_{\mathrm{bf}}, ~~&\text{if}~~ 0\leq K_{\mathrm{bf}}< \bar{K}_{\mathrm{bf}},
  \end{cases}
\end{align}
where $\bar{K}_{\mathrm{bf}}$ is a prescribed upper bound.
\begin{assumption}\label{assump_sat}
It is assumed that the maximum control capability can suppress the disturbance, i.e., $\bar{K}_{\mathrm{bf}}>\bar{d}$.
\end{assumption}
\begin{remark}
Indeed, the saturation function serves to protect a system from an unacceptably large control magnitude. In practice, the upper bound $\bar{K}_{\mathrm{bf}}$ could be specified as the maximum control authority. In case of $\bar{K}_{\mathrm{bf}}<|d(t)|$, an escape of the sliding variable might occur at a certain time moment. However, it would be driven back to the real sliding mode in finite time once $\bar{K}_{\mathrm{bf}}>|d(t)|$. Thus, in contrast to \cite{Obeid 2018}, Assumption \ref{assump_sat} is only sufficient but not necessary.
\end{remark}

\begin{theorem}\label{thr_sat}
Consider the system (\ref{first-order-system}) with the following controller
\begin{align} \label{sat-barrier}
u=-\mathrm{sat}(K_{\mathrm{bf}}(t,s))\mathrm{sign}(s).
\end{align}
If Assumption \ref{assump_sat} holds, for any $s(0)$, the same real sliding surface as in Theorem \ref{theorem new barrier-function adaptive}, $|s(t)|\leq \sigma_1$, is reached in finite time, where $\sigma_1 = \kappa_{\mathrm{bf}}(\bar d)$, $\kappa_{\mathrm{bf}}$ is the inverse of $K_{\mathrm{bf}}$.
\end{theorem}

\begin{IEEEproof}
Consider the Lyapunov function $V=\frac{1}{2}s^2$, whose full time derivative along (\ref{first-order-system}) is given by
\begin{align}
\dot{V}=&s(-\text{sat}(K_{\mathrm{bf}}(t,s))\text{sign}(s)+d)\\ \notag
\leq&-\beta_{\mathrm{sat}}V^{\frac{1}{2}}, \beta_{\mathrm{sat}}=\sqrt{2}(\text{sat}(K_{\mathrm{bf}}(t,s))-\bar{d}).
\end{align}

In the case of $|s(0)|>\sigma_{s}=\kappa_{\mathrm{bf}}(\bar{K}_{\mathrm{bf}})$, where $\kappa_{\mathrm{bf}}$ is the inverse function of $K_{\mathrm{bf}}$, one has $\text{sat}(K_{\mathrm{bf}})=\bar{K}_{\mathrm{bf}}$. By Assumption \ref{assump_sat}, $\beta_{\mathrm{sat}}>0$ holds. This implies the system trajectory reaches the region $|s|\leq \sigma_{s}$ in finite time and the convergence time can be estimated by $t_a\leq \frac{|s_0|-\sigma_s}{\bar{K}_{\mathrm{bf}}-\bar{d}}$.

Once $|s|\leq \sigma_{s}$ is reached, we have $\text{sat}(K_{\mathrm{bf}}(t,s))=K_{\mathrm{bf}}$. By Theorem \ref{theorem new barrier-function adaptive}, the real sliding mode $|s|\leq \sigma_1$ can be reached in finite time. Invoking the mean value theorem for integrals, the convergence time can be estimated by $t_b\leq \frac{\sigma_s-\sigma_1}{K_{\mathrm{bf}}(s^*)-\bar{d}}$, with $s^* \in (\sigma_1,\sigma_s)$. Hence, for any $s_0>\sigma_{s}$, the real sliding mode $|s|\leq \sigma_1$ is reached in finite time and the convergence time satisfies $t_{\mathrm{sat}}\leq t_a+t_b$.

In the case $|s(0)|\leq\sigma_{s}=\kappa_{\mathrm{bf}}(\bar{K}_{\mathrm{bf}})$, the finite time convergence is guaranteed by Theorem $\ref{theorem new barrier-function adaptive}$, and the upper-bound of convergence time is calculated as $t_{\mathrm{sat}}\leq \frac{s_0-\sigma_1}{K_{\mathrm{bf}}(s^*)-\bar{d}}$ with $s^* \in (\sigma_1,s_0)$.
\end{IEEEproof}

\begin{remark}\label{rmk_sat1}
Compared with the non-saturated controller, the accuracy of the sliding mode is unaffected by the introduced saturation function fulfilling Assumption \ref{assump_sat}. However, the convergence time may be greater than that for non-saturated controller.
\end{remark}

\subsection{Barrier function-based adaptive integral sliding mode control with auxiliary dynamics}
The aforegoing control scheme prevents the barrier function-based controller from a large initial control magnitude. However, the overestimation still exits during the reaching phase. Another idea to avoid the large initial control magnitude is eliminating the reaching phase. Following this idea, a barrier function-based adaptive integral sliding mode controller is presented in this subsection. Before elaborating the main design, the following auxiliary dynamics is constructed:
\begin{align}\label{aux_system}
\dot{z}=-\phi(t,z), z(0)=s_0,
\end{align}
where $|\phi(t,z)|\leq L_{\phi}$ is bounded, and $z\phi(t,z)\geq 0$ is assumed, to ensure that the trajectory of (\ref{aux_system}) reaches a small neighbourhood of the origin in finite time $t_z$, i.e., $|z(t)|\leq \sigma_z$ for all $t\geq t_f$.

\begin{theorem}\label{thr_intg}
Consider the system (\ref{first-order-system}) with the following controller:
\begin{align} \label{intg-barrier}
u_{\mathrm{int}}=-\phi(t,z)-K_{\mathrm{bf}}(|e_s|)\mathrm{sign}(e_s),
\end{align}
where $e_s=s-z$. For any $s(0)$, the real sliding surface $|s(t)|\leq \sigma_{\mathrm{int}}=\sigma_1+\sigma_z$, is reached in finite time $t_f$, where $\sigma_{\mathrm{int}}>0$ is tunable.
\end{theorem}

\begin{IEEEproof}
Choose the Lyapunov function as $V_{\text{int}}=\frac{1}{2}e_s^2$, whose full time derivative along (\ref{first-order-system}) with (\ref{intg-barrier}) is calculated as
\begin{align}\label{dv_int}
\dot{V}_{\text{int}}=&e_s(\dot{s}-\dot{z}) \notag\\
\leq& (-K_{\mathrm{bf}}(|e_s|)+d)|e_s|
\end{align}
Since $e_{s}(0)=s(0)-z(0)=0$, by the invariance, $|e_s(t)|\leq \sigma_1$ is guaranteed for any $t\geq t_0$, where $t_0$ is the initial time instant. By $|s|\leq |z|+|e_s|$, $|s(t)|=\sigma_{\mathrm{int}}\leq \sigma_1+\sigma_z$ holds for $t\geq t_f$. If the auxiliary dynamics is designed to achieve the finite-time convergence to the origin, the ultimate bound is reduced to $|s(t)|\leq\sigma_{\mathrm{int}}= \sigma_1$, which is the same as in Theorem \ref{theorem new barrier-function adaptive}. The the control input can be estimated by $|u_{\mathrm{int}}(t)|\leq L_{\phi}+\bar{K}_{\mathrm{bf}}$.
\end{IEEEproof}
\begin{remark}
It can be observed from Theorem \ref{thr_intg} that the convergence property is solely determined by the auxiliary dynamics (\ref{aux_system}) due to the integral sliding mode design which eliminates the transient evolution of $e_s$.
\end{remark}

To enhance Theorem \ref{thr_intg} and design a prescribed-time convergent control law, a time base generator approach is utilized to provide an auxiliary dynamics (\ref{intg-barrier}).
\begin{lemma}\label{lemma_TBG}\cite{Ning 2019}
Consider the differential equation
\begin{align}\label{TBG_SYS}
\dot{p}(t)=-k_g(t)p(t), p(0)=p_0,
\end{align}
with
\begin{align}
k_g(t)=\frac{\dot{\zeta}(t)}{1-\zeta(t)+\delta},
\end{align}
where $\delta>0$ should be chosen sufficiently small, $\zeta(t)$ and $\dot{\zeta}(t)$ satisfy the following properties:
\begin{itemize}
  \item $\zeta(t)$ is at least $\mathcal{C}^2$ on $(0,+\infty)$,
  \item $\zeta(t)$ is continuous and non-decreasing from an initial value $\zeta(t)(0)=0$ to a terminal value $\zeta(t_f)=1$, where $t_f\leq +\infty$ is a prescribed time instant,
  \item $\dot{\zeta}(0)=\dot{\zeta}(t_f)=0$,
  \item $\zeta(t)=1$ and $\dot{\zeta}(t)=0$ for all $t\geq t_f$.
\end{itemize}
Then the system state reaches the value of $\frac{\delta}{1+\delta}p_0$ at time $t_f$, which can be prescribed \emph{a priori} by the user.
\end{lemma}

\begin{theorem}\label{thr_intg_example}
Consider the system (\ref{first-order-system}) with the following controller:
\begin{align} \label{intg-barrier_example}
u_{\mathrm{int}}=-k_g(t)z-K_{\mathrm{cv}}(|e_{s}|)\mathrm{sign}(e_{s}),
\end{align}
where $e_s=s-z$ and $z$ is generated by the following auxiliary dynamics:
\begin{align}\label{TBG_example}
\dot{z}=-k_g(t)z, z(0)=s_0;
\end{align}
where
\begin{align}\label{s_function_1}
\zeta(t)=\begin{cases}
             \frac{1-\cos(\frac{\pi t}{t_f})}{2}, & \mbox{if } t\leq t_f,\\
             1, & \mbox{otherwise},
           \end{cases}
\end{align}
and
\begin{align}\label{ds_function_1}
\dot{\zeta}(t)=\begin{cases}
             \frac{\pi \sin(\frac{\pi t}{t_f})}{2t_f}, & \mbox{if } t\leq t_f,\\
             0, & \mbox{otherwise}.
           \end{cases}
\end{align}
For any $s(0)$, the real sliding surface $|s(t)|=\sigma_{\mathrm{tbg}}\leq \frac{\delta}{1+\delta}s_0+\sigma_{\mathrm{cv}}$, is reached within the prescribed time $t_f$.
\end{theorem}
\begin{IEEEproof}
In line with the proof of Theorem \ref{thr_intg}, $e_{s}(0)=0$ guarantees that the real sliding mode $|e_{s}|\leq \sigma_{\mathrm{cv}}=\kappa_{\mathrm{cv}}(\bar{d})$ is reached for all $t\geq t_0$. Then, $|s(t)|\leq |z(t)|+\sigma_{\mathrm{cv}}$ follows for any $t\geq t_0$. Since (\ref{s_function_1}) and (\ref{ds_function_1}) satisfy the conditions of Lemma \ref{lemma_TBG}, $|z(t)|=\frac{\delta}{1+\delta}s_0$ is achieved for all $t\geq t_f$. Hence, the finite-time convergence to the real sliding surface $|s(t)|\leq \sigma_{\mathrm{tbg}}=\frac{\delta}{1+\delta}s_0+\sigma_{\mathrm{cv}}$ is ensured for all $t>t_f$.
\end{IEEEproof}
\begin{remark}
The proposed controller (\ref{intg-barrier_example}) presents the following benefits: 1) the initial control magnitude is zero; 2) the control input is chattering-reduced and bounded; 3) the convergence time can be prescribed.
\end{remark}

\begin{remark}
The two control schemes proposed in this section enable one to avoid large control magnitudes and guarantee global stability. Furthermore, the designed controllers are continuous, in contrast to the one given in \cite{Obeid 2018}.
\end{remark}

\section{Simulations}\label{Sec sim}
\begin{figure}
  \centering
  \includegraphics[scale=0.5]{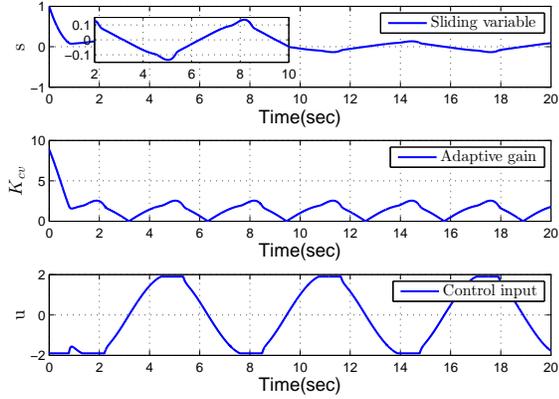}
  \caption{Responses to controller (\ref{cv-barrier}) in presence of input saturation}\label{Saturate new_barr}
\end{figure}

Recall the claim in Remark \ref{remark motivation} that the proposed control methodology can relax the requirement that the maximum control magnitude should be superior to the disturbance bound at every time moment. To validate this claim, the system (\ref{first-order-system}) with the same disturbance and initial condition as in Section \ref{Sec Motivation} is considered. The barrier function (\ref{cv-barrier}) with $\rho_1=5$ and $\lambda_1=0.2033$ is chosen to achieve the same accuracy $|s|\leq 0.1$ for comparison. The control input is also constrained by $|u|\leq 1.9$. In contrast to Fig. \ref{Saturate impact}, the sliding variable in Fig. \ref{Saturate new_barr} presents only a transient escape but returns to the real sliding mode when the disturbance fades away, which is consistent with the claim in Remark \ref{remark motivation}.

\begin{figure}
  \centering
  \includegraphics[scale=0.45]{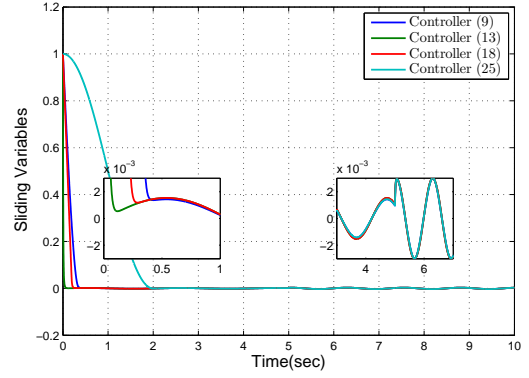}
  \caption{Time histories of sliding variables}\label{fig-sliding_variabel__All}
\end{figure}

\begin{figure}
  \centering
  \includegraphics[scale=0.45]{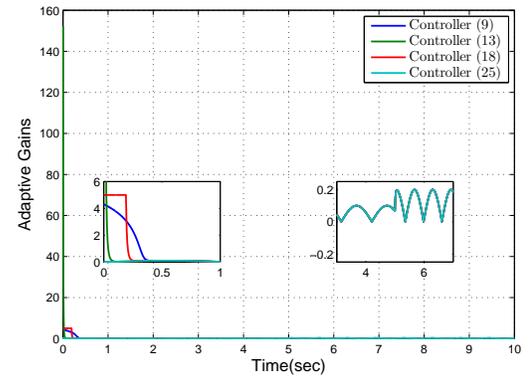}
  \caption{Time histories of adaptive gains}\label{fig-adaptive gain_All}
\end{figure}

\begin{figure}
  \centering
  \includegraphics[scale=0.45]{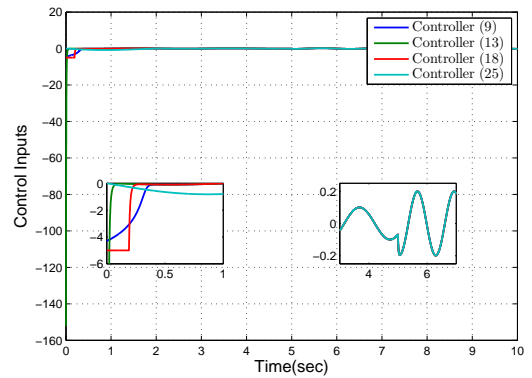}
  \caption{Time histories of control inputs}\label{fig-control_input__All}
\end{figure}

To demonstrate effectiveness of the proposed controllers, consider the system (\ref{first-order-system}) with the disturbance
\begin{align}\label{disturbance}
d(t)=\begin{cases}
       0.1\sin(3t), & \mbox{if } t\leq 5~\mathrm{s}, \\
       0.2\cos(5t), & \mbox{otherwise},
     \end{cases}
\end{align}
from which the upper bound $\bar d=0.2$ is established. The parameters in controllers (\ref{cc-barrier}), (\ref{cv-barrier}), (\ref{sat-barrier}) and (\ref{intg-barrier_example}) are set to $\rho_1=1$, $\lambda_1=0.01366$, $\rho_2=5$, $\lambda_2=0.05$, $\bar{K}_{\mathrm{bf}}=5$, $t_f=2$, and $\delta=10^{-5}$, respectively. Thus, by (\ref{kappa_cc_d}) and (\ref{kappa_cv_d}), the same accuracy $\sigma_{\mathrm{cc}}=\sigma_{\mathrm{cv}}\approx 3.02\times 10^{-3}$ is guaranteed for closed-loop performance comparisons. The same initial value of the sliding variable $s_0=1$ is assigned. The sampling time is set to $h=0.001 \mathrm{s}$. All closed-loop responses under the controllers proposed in this paper are shown in Figs. \ref{fig-sliding_variabel__All}-\ref{fig-control_input__All}. From Fig. \ref{fig-sliding_variabel__All}, it can be observed that the sliding variables in (\ref{first-order-system}) with the proposed controllers reach the same real sliding mode $|s|\leq 3.02\times 10^{-3}$, and the convergence time corresponding to the controller (\ref{cv-barrier}) is less than the one corresponding to the controller (\ref{cc-barrier}) or (\ref{sat-barrier}), which is consistent with Proposition \ref{prop comp} and Remark \ref{rmk_sat1}. Particularly, the sliding variable under the controller (\ref{intg-barrier_example}) reaches the real sliding surface $|s|\leq \sigma_{\mathrm{cv}}$ at $t=2$ seconds, which is a prescribed convergence time. The time histories of the adaptive gains and control inputs are depicted in Fig. \ref{fig-adaptive gain_All} and Fig. \ref{fig-control_input__All}, which show that there is no gain overestimation after the real sliding mode is established. Thus, the simulation results demonstrate effectiveness of the class $\mathcal{K}_\infty$ function-based controllers as well as their reliable performance.

\section{Conclusion} \label{Sec conclusion}
This paper presents a class $\mathcal{K}_\infty$ function-based adaptive sliding mode control methodology for systems with matched disturbances. To demonstrate this new idea, two different barrier functions are introduced, both of which are special instances belonging to the class $\mathcal{K}_\infty$. The proposed barrier function-based adaptive sliding mode controllers guarantee that the real sliding mode is achieved in finite time. Finally, the given simulation results and discussions demonstrate effectiveness and merits of the proposed control design.

\end{document}